\documentstyle[12pt,aps,prc,epsfig]{revtex}
\tightenlines
\begin{document}
\begin{titlepage}

\title{Phenomenology of $K^-$ atoms and other strange hadronic atoms}
\author{A. Gal$^{a}$\\ $^{a}$Racah Institute of Physics, The Hebrew
University, Jerusalem 91904, Israel}
\maketitle

\begin{abstract}
\newline
\newline
Recent optical-potential studies of the phenomenology of $K^-$ atoms
are reviewed. It is shown that the data can be fitted by a complex
optical potential with either a relatively shallow attractive
component ($\sim -50$ MeV at $\rho_0$), as suggested by a self
consistent application of chirally motivated coupled-channels models,
or a relatively deep attractive component ($\sim -180$ MeV at $\rho_0$)
as suggested by a mean-field extrapolation of a phenomenological
low-density expansion. Both classes of these optical potentials,
due to their strongly absorptive component, predict similarly structured
relatively narrow `deeply bound' $K^- \it atomic$ states with widths
saturating at about 2 MeV. Possible formation reactions are briefly 
discussed. The more speculative case for relatively
narrow deeply bound $K^- \it nuclear$ states is briefly mentioned.
Recent works on other strange hadronic atom systems, for the $\Sigma^-$
and $\Xi^-$ hyperons, are also discussed.
\newline
\newline
Invited talk presented at HYP2000 in Torino, October 2000. To appear in
Nuclear Physics A.
\end{abstract}
\vspace{5cm}
\centerline{\today}
\end{titlepage}

\section{KAONIC ATOMS}
\label{sec:kaons}

In this section I review recent works on the $K^-$ nucleus
optical potential, with emphasis in the first two subsections on 
fitting it to the `normal' X-ray data in kaonic atoms, 
and in the third subsection on predictions of narrow `deeply bound' 
$K^-$ atomic states.

\subsection{Hybrid relativistic mean field approach to $K^-$ atoms}

Here I report on the recent work by the Jerusalem-\v{R}e\v{z}
collaboration \cite{FGM99}. Antikaons are incorporated into
this relativistic mean field (RMF) model by using the Lagrangian
density of the form \cite{sm96}
\begin{equation}\label{equ:lagrange}
{\cal L}_{K} = \partial_{\mu}\overline{\psi}\partial^{\mu}\psi -
m^2_K\overline{\psi}\psi
- g_{\sigma K}m_K\overline{\psi}\psi\sigma
 - ig_{\omega K}(\overline{\psi}\partial_{\mu}\psi {\omega}^{\mu} -
\psi \partial_{\mu}
\overline{\psi}{\omega}^{\mu})
+(g_{\omega K}{\omega}_{\mu}
)^2
\overline{\psi}\psi,
\end{equation}
describing the interaction of the antikaon field
($\bar{\psi}$) with the scalar ($\sigma$) and vector ($\omega$)
isoscalar fields.
The corresponding equation of motion for $K^-$ in a $Z = N$ nucleus can be
expressed by a Klein-Gordon equation with the real part of the
optical potential given at threshold by

\begin{equation}\label{equ:VOP1}
{\rm Re }\;V_{\rm opt}={{m_K}\over{\mu}}({1\over{2}}S - V - {{V^2}\over{2m_K}})
\;\;\; ,
\end{equation}
where $S = g_{\sigma K}\sigma(r)$ and $V = g_{\omega K}\omega_0(r)$ in terms
of the mean isoscalar fields, and where $\mu$ is the $K^-$ nucleus
reduced mass.
Note that for antikaons, the vector potential $V$ contributes attraction,
just opposite to its role for kaons and nucleons. Each of the three
terms on the r.h.s. of Eq. (\ref{equ:VOP1}), thus,
gives rise to attraction. Consequently,
it becomes impossible to satisfy the low density limit which requires
that, due to the subthreshold $\Lambda$(1405), Re~$V_{\rm opt} > 0$ as
$\rho \rightarrow 0$.
For nuclei with $N>Z$, the potential should include also an isovector
part due to the interaction of the $K^-$ with the $\rho$ meson field.
However, this was omitted from the present calculations as
it was found in previous analyses of kaonic atoms \cite{fgb93,fgb94,bfg97}
to have a marginal effect. Furthermore, since the imaginary part of
$V_{\rm opt}$ is not directly addressed in the RMF approach, a phenomenological
$t\rho$ parameterization was used for it, after making sure that a more
involved density dependence of Im~$V_{\rm opt}$ does not introduce further
effects.

In order to construct the RMF Re~$V_{\rm opt}$ of Eq.~(\ref{equ:VOP1}), 
one needs to specify $\alpha_{\sigma}$ and $\alpha_{\omega}$, 
where $\alpha_m = g_{mK}/g_{mN}$. Treating both $\alpha_{\sigma}$ and 
$\alpha_{\omega}$ as free parameters, the best fit potential consists of 
a strongly attractive vector potential and a strongly {\it repulsive} 
scalar potential, quite far from the structure expected e.g. from the 
naive quark model (QM) \cite{br96}. The corresponding Re~$V_{\rm opt}$ 
provides a very good fit to the atomic data. It is attractive in the 
nuclear interior, reaching a depth of about 190~MeV in close agreement 
with the phenomenological density-dependent (DD) potentials of Ref. 
\cite{bfg97}, and it becomes repulsive at large distances, thus reflecting
{\it a posteriori} effects of the $\Lambda$(1405) at low densities.

The RMF description is well justified within the nuclear interior, for
densities larger than about 0.2$\rho_0$ where the effects of the 
$\Lambda$(1405) may be neglected, as demonstrated in 
Refs.~\cite{k94,wkw96,or98}. The existence of the $\Lambda$(1405) 
resonance clearly poses a difficulty for the RMF approach at low densities 
if the parameters $\alpha_m$ are allowed to deviate only moderately 
from the values suggested by the underlying hadron symmetries. In the
hybrid model of Ref. \cite{FGM99}, the functional RMF form
Eq.~(\ref{equ:VOP1}) was used in the nuclear interior for Re~$V_{\rm opt}$,
whereas a purely phenomenological DD form \cite{fgb93,fgb94} 
was used in the surface of the
nucleus and beyond. The sensitivity to the choice of the radius $R_M$,
where the two forms are matched to each other, was checked and found
to be small. The density $\rho(R_M)$ is sufficiently high to justify
using the RMF approach, and sufficiently low so that the atomic data
are still sensitive to the RMF form. Figure~4 of Ref.~\cite{bfg97},
and in particular Fig.~3 of Ref.~\cite{f98} for Ni, show that by analyzing
kaonic atoms, using a given functional form for $V_{\rm opt}$, one determines
the real part of the $K^-$ nucleus DD optical potential up to
$\rho=0.9\rho_0$. This is well above the density at which the RMF
form takes over in the present approach.

\begin{figure}
\begin{minipage}[t]{80mm}
\epsfig{file=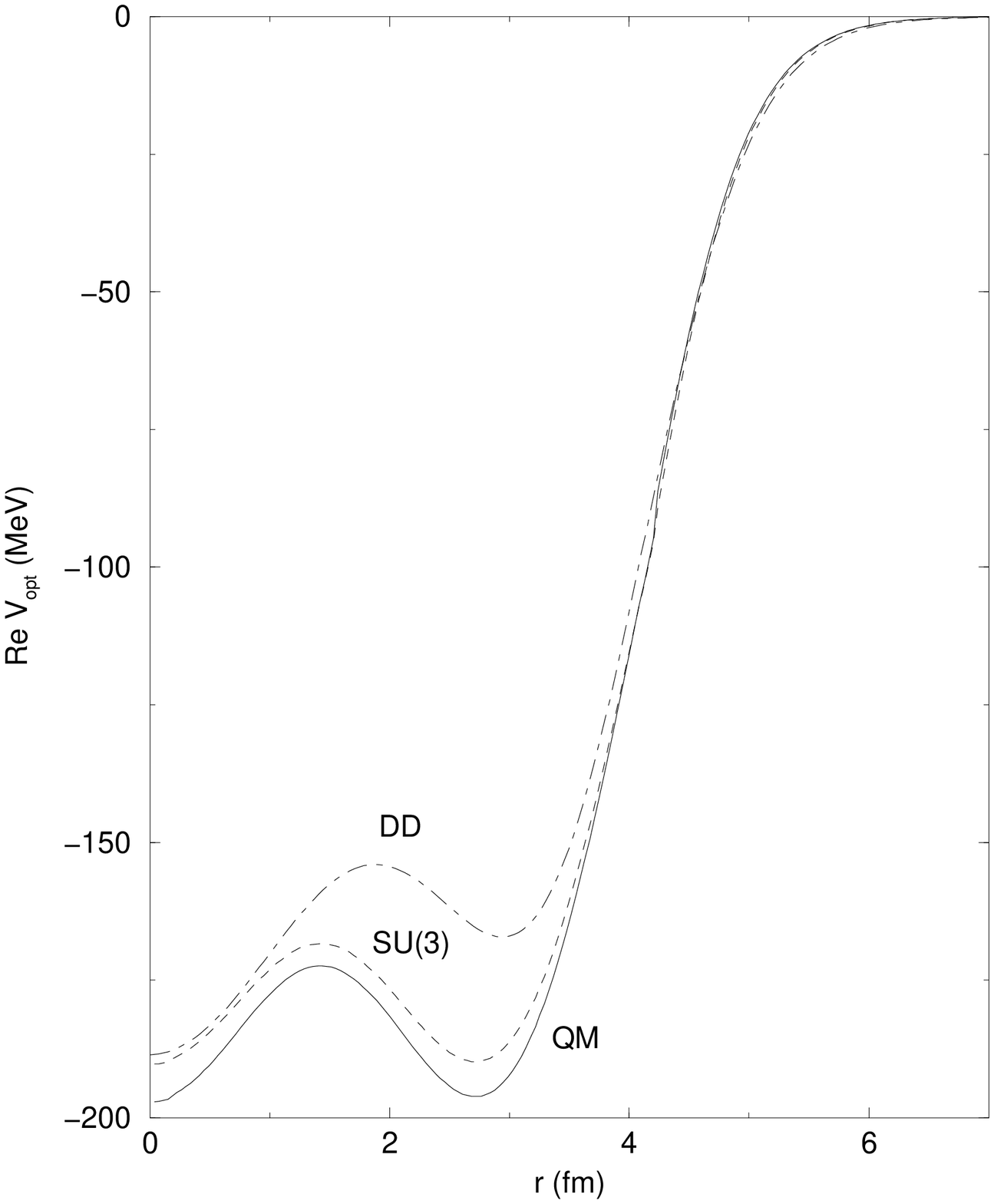,height=83mm,width=80mm,
bbllx=31,bblly=92,bburx=509,bbury=670}
\caption{Combined RMF+DD best-fit real potentials for $K^-$ in Ni, using 
the QM and SU(3) versions. Also shown is the phenomenological DD potential.}
\label{fig:NiPb}
\end{minipage}
\hspace{\fill}
\begin{minipage}[t]{75mm}
\epsfig{file=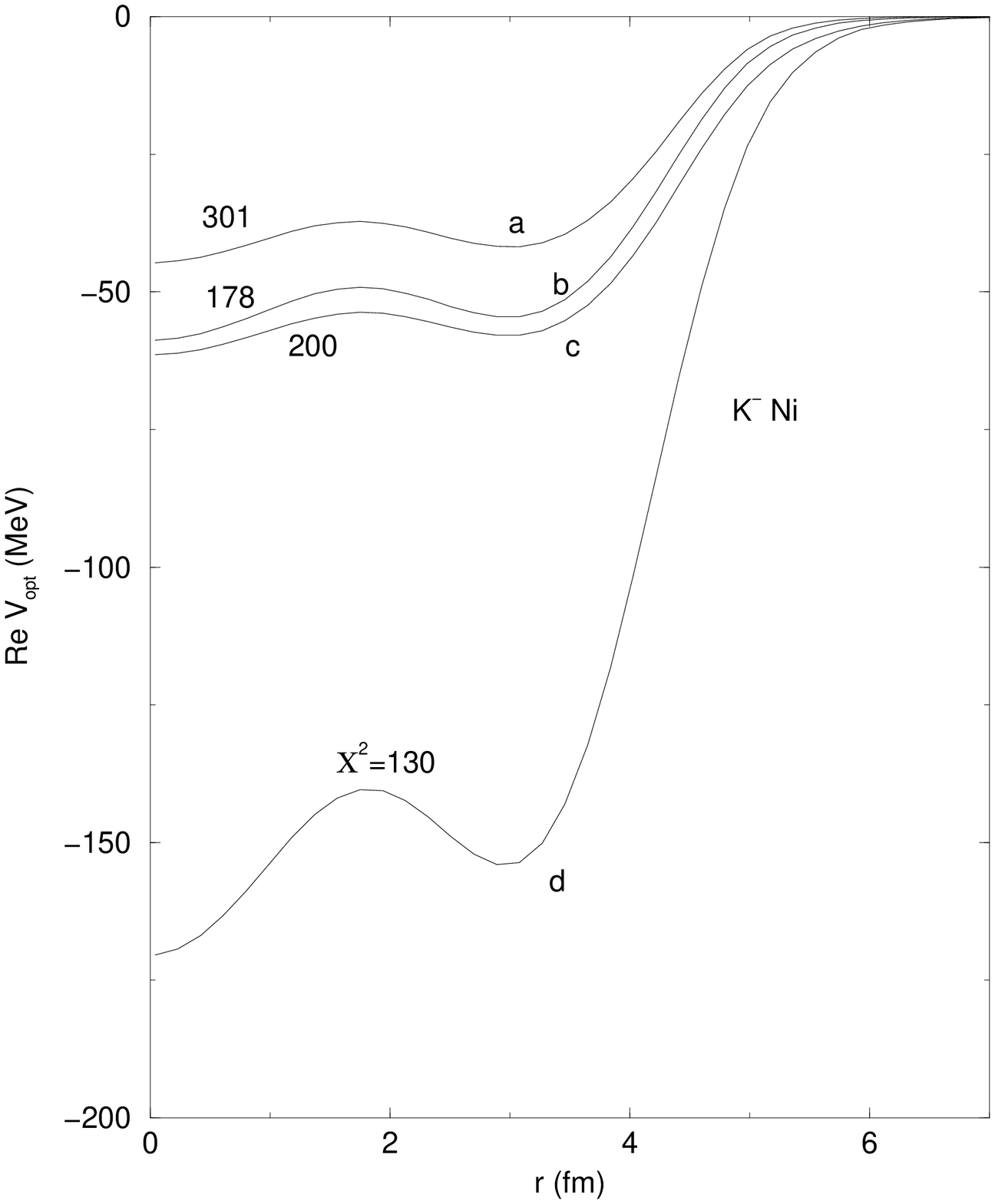,height=83mm,width=75mm,
bbllx=31,bblly=91,bburx=509,bbury=670}
\caption{Best-fit real potentials, (b)-(d), for $K^-$ in Ni using the
self consistent potential (a) due to Ramos and Oset. The values of total
$\chi^2$ are given for each curve.}
\label{fig:ramos}
\end{minipage}
\end{figure}

Fitting to the $K^-$ atomic data, the vector coupling constant ratio
$\alpha_{\omega}$ was kept fixed, guided by theoretical considerations,
and the scalar coupling constant ratio $\alpha_{\sigma}$
was varied together with the parameters of the DD real potential form 
(for $r>R_M$). For the coupling constant $g_{\omega K}$, either the
constituent QM value $\alpha_{\omega}={1 \over 3}$ was used,
or the SU(3) relation $2g_{\omega K}=g_{\rho \pi}=6.04$.
The resulting scalar potentials are attractive, of the order
of magnitude expected (see Table I in Ref. \cite{FGM99}).
These best-fit hybrid RMF+DD potentials describe the data very well, 
with $\chi^2/N$ values of $1.4 - 1.5$.

Figure \ref{fig:NiPb} shows the hybrid RMF+DD best fit real potentials
of Ref. \cite{FGM99} for Ni, using the QM and SU(3) options. It is seen that
the depths of the hybrid real potentials in the nuclear interior
for the different choices of $\alpha_{\omega}$ are about 180~MeV, 
very close to the depth of the purely phenomenological DD potential, also 
shown in the figure. These depths are to be compared to those 
displayed in Fig. \ref{fig:ramos} discussed below. 
We note that omitting the $V^2$ term
in Eq.~(\ref{equ:VOP1}), and refitting the $K^-$ atomic data, the resulting
depths are smaller than the RMF+DD depths shown in Fig. \ref{fig:NiPb}
by less than 10 MeV.
Finally, as for Im~$V_{\rm opt}$, it is well determined in all
of the above best fits, with a depth of about 60 MeV, not far from the value
expected from the low density limit.

\subsection{Self consistent optical potentials in chirally inspired models}

In this approach (reviewed by Ramos in these proceedings) 
chirally motivated coupled-channel $t$ matrices
in the $S=-1$ strangeness sector, including the ${\bar K}N$ channels,
were fitted to the near-threshold data and used to construct
a ${\tilde t}\rho$ $K^-$ nucleus optical potential \cite{wkw96,or98}.
Here $\tilde t$ is a medium-modified ${\bar K}N$ $t$ matrix incorporating
Pauli blocking in the intermediate nucleon states, nucleon and hyperon
dispersive corrections and also, recently, self consistency
\cite{Lutz98,ro00} (and \cite{SKE00} within a non chiral model).
This latter requirement of self consistency means that the output 
$V_{\rm opt}$ is used in the $\bar K$ nuclear-medium propagator
within the Lippmann Schwinger equation to generate
the $\tilde t$ matrix, and therefore also the input $V_{\rm opt}$,
from the coupled-channel interaction $v$. A common feature of the self
consistent evaluations is that Re $V_{\rm opt}$ is rather shallow,
typically about $-40$ MeV at $\rho_{0}$, as shown by curve ($a$) 
in Fig. \ref{fig:ramos}.

The $K^{-}$ nucleus optical potential derived self consistently by
Ramos and Oset \cite{ro00} has been applied in Ref. \cite{HOT00} to a
subset of the $K^{-}$ atomic data (including the same $2p$, $3d$ and $4f$
sequences already covered in Ref. \cite{mht94}) which it reproduces
semi-quantitatively. Baca et al. \cite{BGN00} have recently extended
this analysis, including the missing sequences for heavier atoms, and
also allowing for an additional phenomenological component of a
$(\delta t) \rho$ form. Figure \ref{fig:ramos}, taken from a work in
progress \cite{CFG01}, shows the successively decreasing total $\chi^2$ 
values for $N=65$ data points,
plus the calculated depths of Re $V_{\rm opt}$, when such addition is
allowed (case $b$). The potential has become somewhat more attractive,
reaching about $-60$ MeV at $\rho_{0}$. Shown also are two more
searches, starting with the Ramos and Oset $V_{\rm opt}$ (case $a$). In
the first case ($c$), the $I = 0$ component which is intimately
related to the $\Lambda(1405)$ propagation and dissolution in the medium 
is kept as is, whereas the strength of the $I = 1$ component which is
not constrained well by the low-energy $\bar KN$ data is allowed to get 
renormalized, while keeping its
`theoretical' shape intact. This kind of a search is somewhat less
successful than searching on a $(\delta t) \rho$ form, yielding 
a similar depth for Re $V_{\rm opt}$. 
Trying a combination of the above two modifications
(case $d$) reduces the value of $\chi^2$ considerably, 
close to that due to the completely phenomenological DD analysis 
\cite{bfg97}, producing a very deep potential (about $-170$ MeV at 
$\rho_{0}$). Similar results are obtained if one starts from the 
self consistent $K^-$ nucleus potential of Ref. \cite{SKE00}.

\subsection{Narrow deeply bound $K^-$ and $\bar p$ atomic states}

Friedman and Soff \cite{FSo85} and Toki and Yamazaki \cite{TYa88}
predicted that $\pi^-$ atomic levels
remain relatively narrow and well resolved, with absorption widths
smaller than 1 MeV, down to the $1s$ level in Pb.
This narrowness is due to the
repulsive $s$-wave pion interaction which `pushes' its
wave function outside of the nucleus,
thus suppressing the absorption width. Recently the $1s$ and
$2p$ $\pi^-$ `deeply bound' atomic states (DBAS), which are inaccessible
via the atomic cascade process, have been observed at GSI \cite{Gil99}
using the recoilless ($d,^{3}$He) reaction on two Pb isotopes,
confirming the theoretical prediction.

For $K^-$ and $\bar p$ atoms, Re $V_{{\rm opt}}$ is strongly attractive
and, furthermore, Im $V_{{\rm opt}}$ is particularly strong
(50 - 100 MeV deep \cite{bfg97}), so DBAS a-priori are unlikely to
be narrow. However, as shown recently,
a strongly absorptive potential makes $V_{{\rm opt}}$ {\it effectively
repulsive}, and the atomic wave functions are then substantially
suppressed within the nucleus, with the atomic level
widths saturating as function of Im $V_{{\rm opt}}$. Figures 3-6 below
demonstrate these points \cite{FGa99a,FGa99b}.

\begin{figure}
\begin{minipage}[t]{80mm}
\epsfig{file=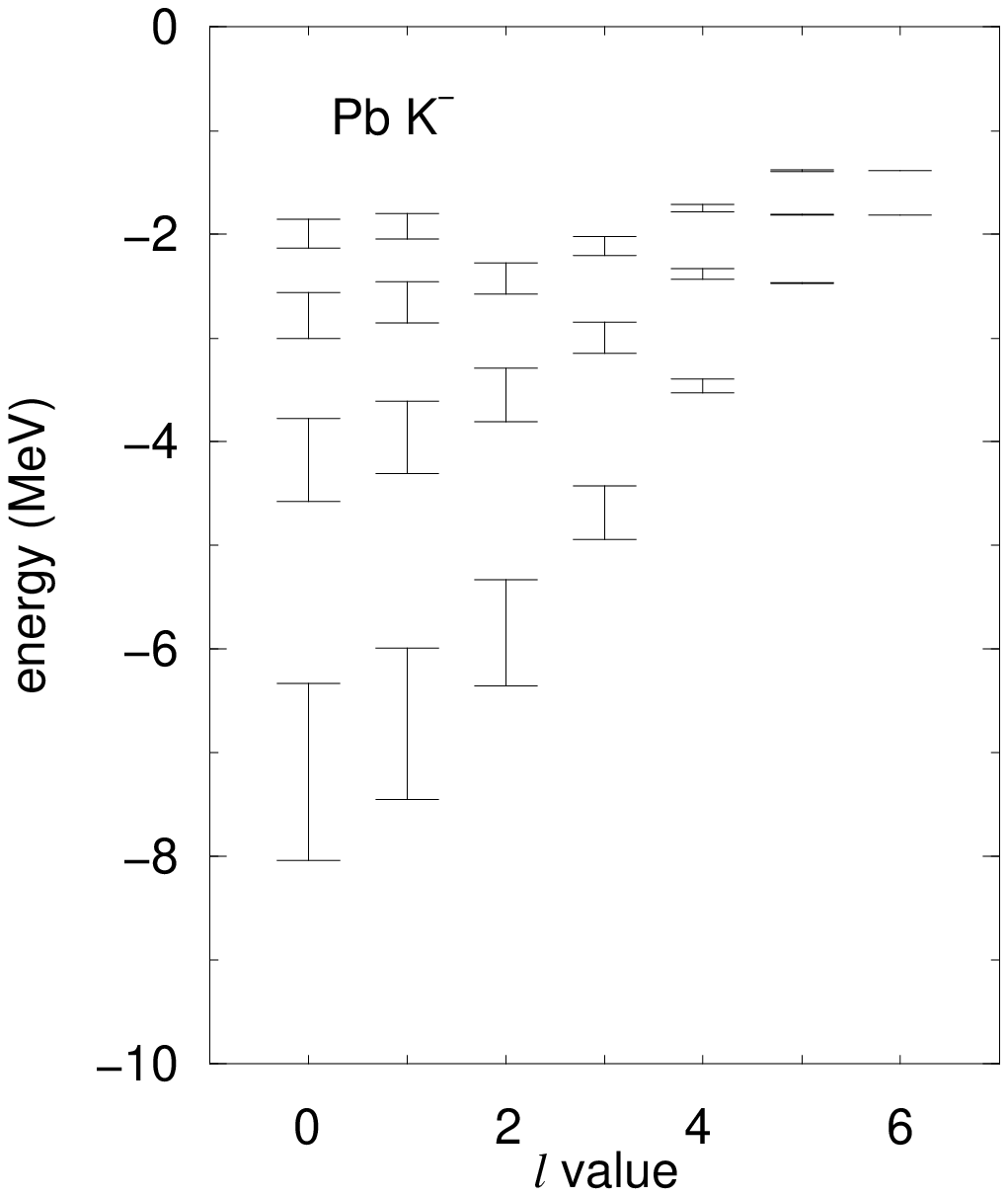,height=87mm,width=80mm,
bbllx=129,bblly=205,bburx=422,bbury=556}
\caption{Energy levels for $K^-$ atomic Pb.}
\label{fig:PbspectK}
\end{minipage}
\hspace{\fill}
\begin{minipage}[t]{75mm}
\epsfig{file=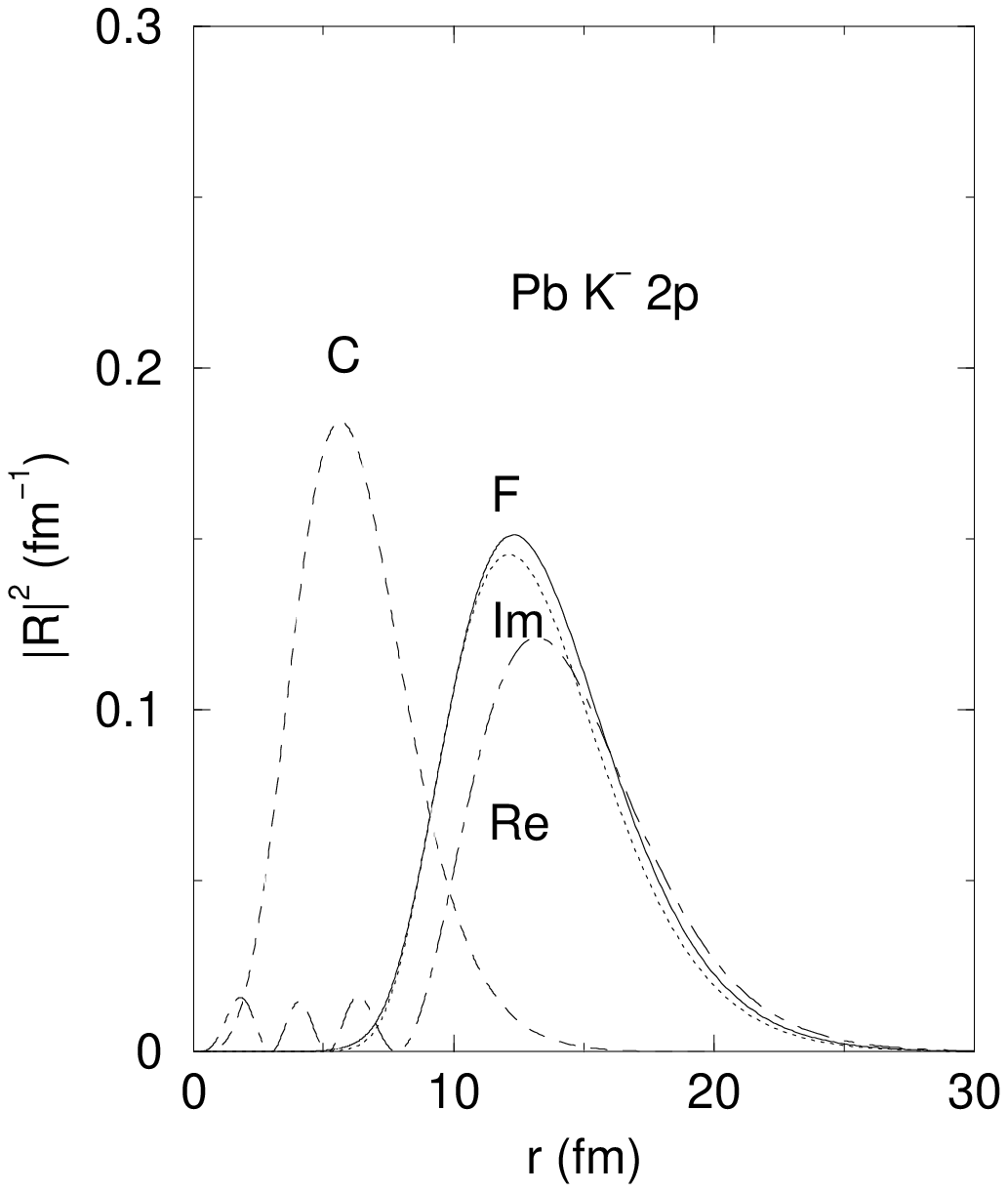,height=87mm,width=75mm,
bbllx=134,bblly=205,bburx=430,bbury=556}
\caption{Squared $2p$ radial wavefunctions in $K^-$
atomic Pb.}
\label{fig:KPb2pwf2}
\end{minipage}
\end{figure}

Figure \ref{fig:PbspectK} shows calculated energy levels for kaonic Pb.
The bars stand for the full width $\Gamma$ of the level. The $7i$ $K^-$
level in Pb is the last observed in the X-ray spectrum.
Clearly the deeper levels are quite well defined even for a heavy nucleus,
but $l$-selective reactions are needed to observe these DBAS. These 
predictions are insensitive to details of $V_{\rm opt}$, provided it was
fitted to the `normal' levels derived from X-ray spectra.

Figure \ref{fig:KPb2pwf2} shows the moduli squared of 2$p$
radial wave functions in kaonic Pb.
The dashed curve (C) is for the finite size Coulomb potential
only. The solid curve (F) is for the full $V_{{\rm opt}}$
added and it shows essentially total expulsion of the wave function from
the nucleus, whose r.m.s. radius is 5.5 fm,
thus reducing dramatically the width of the level. The dotted curve (Im)
is when only Im $V_{{\rm opt}}$ is added, showing that
it is dominant in determining the wave function.
The dot-dashed curve (Re) shows the wave function when only the
{\it strongly attractive}
Re $V_{{\rm opt}}$ is added.
The three small inner peaks preceding the main peak well outside the
nucleus indicate that three
nuclear bound states exist in this real potential, causing the atomic
wave function (by orthogonality)
to develop nodes. This inner structure then causes the main
{\it atomic} peak of the wave function to shift to larger radii
compared to the Coulomb wave function, thus resulting
in a {\it repulsive shift}.
The node structure inside the nucleus depends critically on Re $V_{{\rm opt}}$
and so are the energies of the {\it broad nuclear} bound states.
The atomic part of the wave function is much less sensitive.

\begin{figure}
\begin{minipage}[t]{80mm}
\epsfig{file=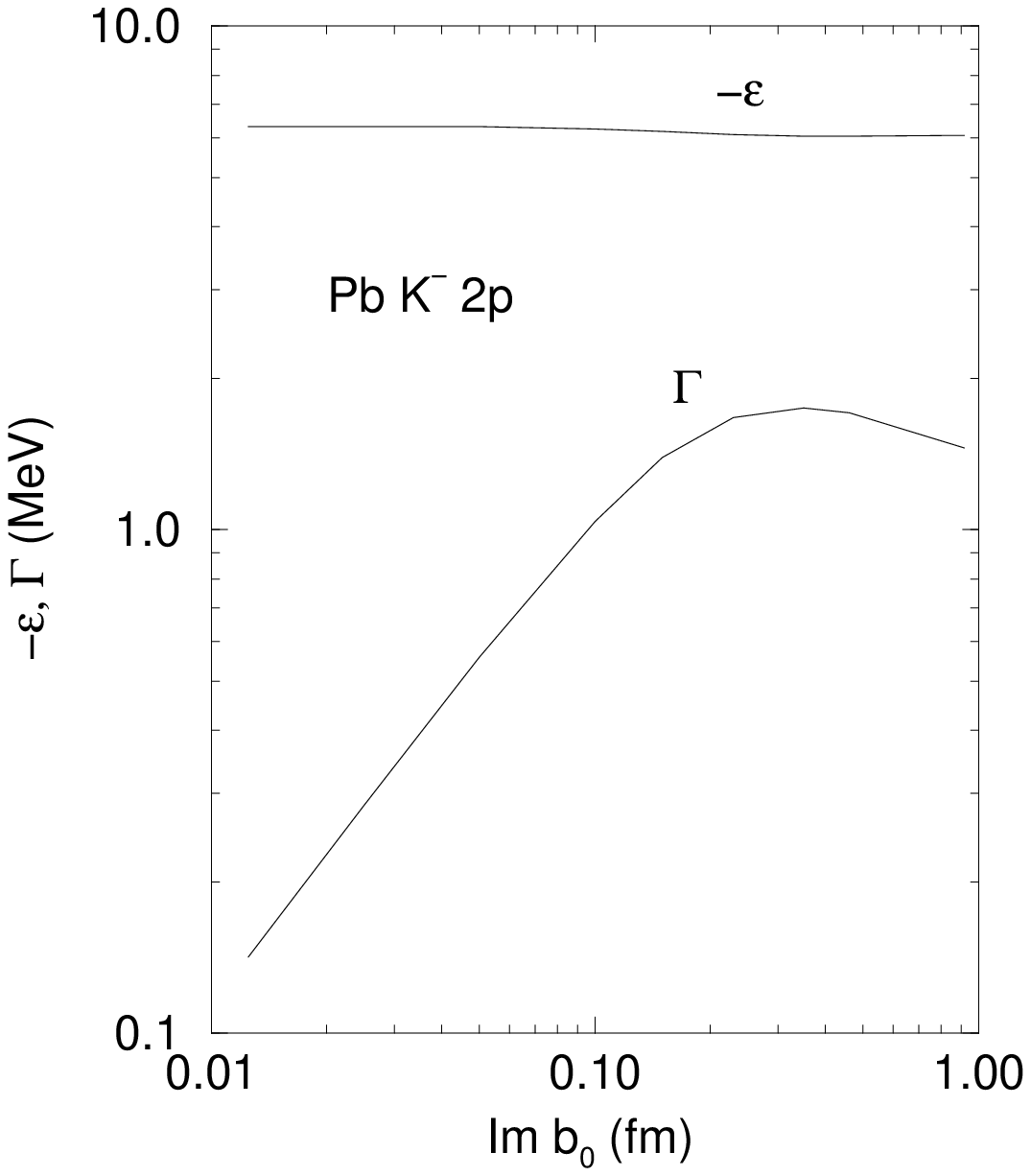,height=87mm,width=80mm,
bbllx=125,bblly=202,bburx=436,bbury=556}
\caption{Strong interaction shifts and widths
as function of Im $b_0$ for the $2p$ level in kaonic Pb.
Re~$b_0$=0.62~fm.}
\label{fig:KPb2peg}
\end{minipage}
\hspace{\fill}
\begin{minipage}[t]{75mm}
\epsfig{file=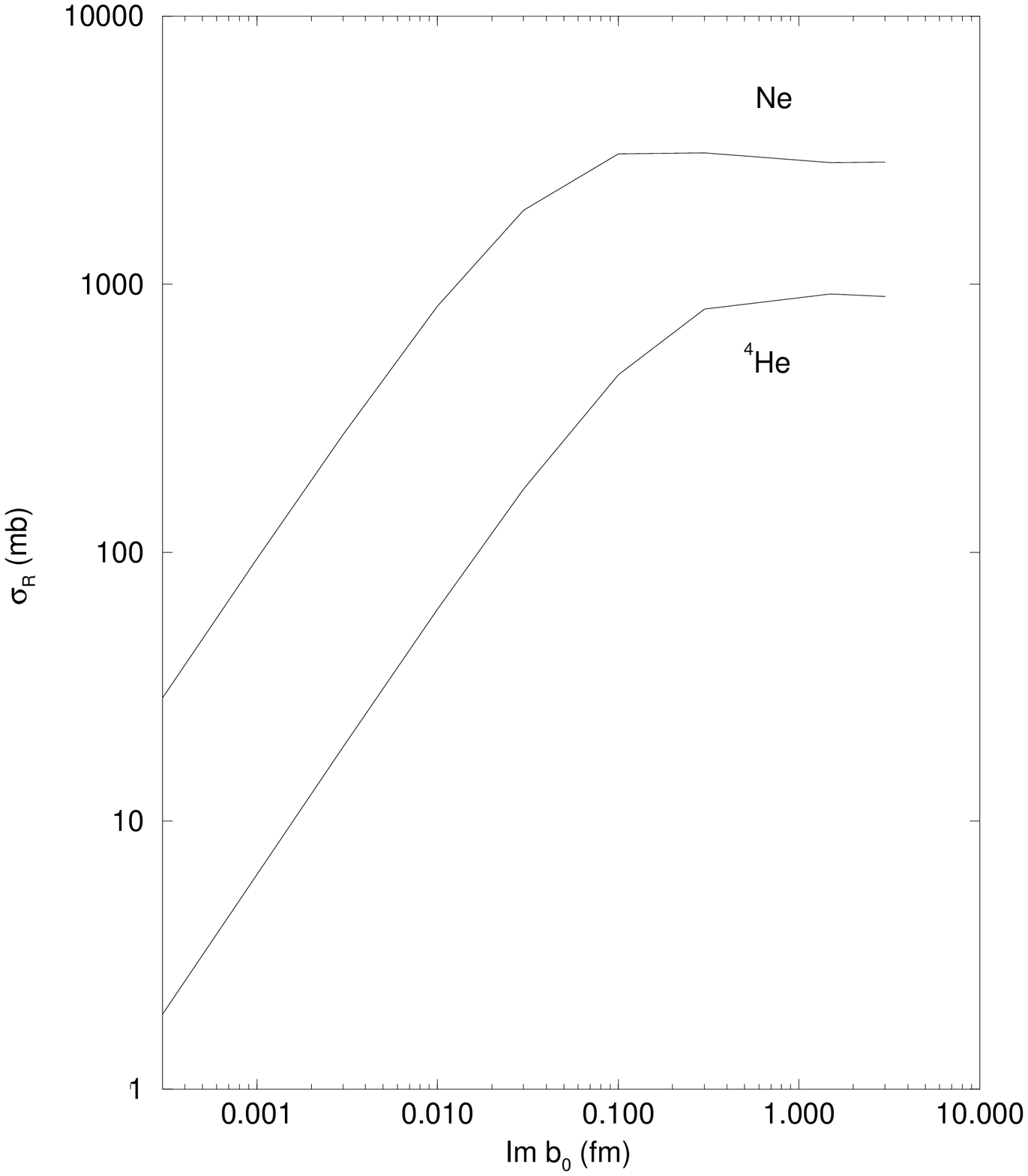,height=87mm,width=75mm,
bbllx=22,bblly=89,bburx=530,bbury=670}
\caption{Total reaction cross sections for 57 MeV/c $\bar p$ on
$^4$~He and Ne as function of Im~$b_0$.}
\label{fig:sat}
\end{minipage}
\end{figure}

The saturation of atomic level widths may be demonstrated by studying the
dependence of strong interaction level shifts $\epsilon$ and widths
$\Gamma$ on Im $V_{{\rm opt}}$.
Figure \ref{fig:KPb2peg} shows calculated shifts and widths for the $2p$
level in kaonic Pb as function of Im $b_0$
when Re $b_0$ is being held at its nominal
value, using a $t\rho$ potential with effective scattering length $b_0$.
It is seen that at about 20\% of the nominal value of Im $b_0$ (of 0.92 fm)
the width already saturates and then it goes slowly down, while the
shift stays essentially constant at a very large repulsive value,
in spite of the strongly {\it attractive} Re $V_{{\rm opt}}$.
The mechanism behind the saturation of widths is fairly independent of
Re $V_{{\rm opt}}$, as demonstrated in the previous figure. In the absence
of data on this part of the atomic spectrum, the validity of the
saturation property is supported by studying total reaction cross
sections at very low energies. Related data do exist for $\bar p$
annihilation on a few nuclear targets. Figure \ref{fig:sat} shows
calculated total reaction cross sections for $\bar p$ at 57 MeV/c
on $^4$He and Ne as function of Im $b_0$ \cite{GFB00,BFG00}. The
measured cross sections (not shown in the figure) are well reproduced
by the plateau values. The striking similarity between Fig. \ref{fig:KPb2peg}
and Fig. \ref{fig:sat} is due to the observation that both $\Gamma$
at negative energies, and $\sigma _{R}$ at positive energies, involve 
averaging Im $V_{\rm opt}(r)$ with $|\psi|^{2}$.
Therefore, in the strong-absorption limit, when $|\psi|^{2}$ is
pushed out of the nuclear domain (as shown in Fig. \ref{fig:KPb2pwf2})
these observables cease to rise approximately linearly with the
strength of Im $V_{\rm opt}$.

The lesson gained from discovering $\pi^-$ DBAS \cite{Gil99} 
is that nuclear reactions with momentum transfer $q$ below 50 MeV/c 
should suit best the production of $K^-$ DBAS. First, 
the suppressive effects of the nuclear absorption on low-$q$ 
production cross section are weaker than for
reactions with higher values of $q$ \cite{NOs90} and second, the
angular momentum transfer is then minimal: ${\Delta}l \sim 0$,
thus ensuring {\it selectivity}. 

Low energy (16 MeV) $K^-$ mesons are now available at
DA$\Phi$NE due to the decay at rest  $\phi (1020)~\rightarrow~
K^{+} K^{-}$.
This corresponds to $q(0^o) = 47$ MeV/c in the ($K^{-}, p$) reaction.
Although forward cross sections as large as 0.1 $\mu$b/sr may be
expected, a major problem would be the need to detect unambiguously
the low energy outgoing proton. Somewhat smaller cross sections are
expected for the $(K^{-}, \gamma)$ reaction \cite{HOT00} in which
$q(0^{o})$ is about twice as large. Here too, the expected background
at these low energies poses a serious problem. If the decay at rest of the
$\phi (1020)$ meson were
to take place at close proximity to a nucleus such that the kaonic
atom recoiled as a whole, then the signature would be
a peak in the $K^+$ energy spectrum above the 32 MeV released in the
free-space $\phi$ decay.
However, since $q \sim $ 180 MeV/c in this at-rest $(\phi, K^{+})$
reaction, it is not a favorable situation.
It would be interesting to look for secondary nuclear interactions
of the $\phi$ meson {\it in flight} prior to its decay, e.g. the
recoilless $(\phi, K^{+})$ reaction which attaches
a $K^{-}$ meson to the nuclear target \cite{FGa99b}.

Several authors \cite{Kis99,AYa99} have drawn attention to $K^{-}$
{\it nuclear} states bound by a very deep potential, similar to the
best-fit DD potential of Batty et al. \cite{bfg97}. The deepest
states are then blocked from decaying by the two-body mode $\bar KN
\to \pi \Sigma$ and their width could then be
reduced to 10 MeV. However, about 20\% of $K^{-}$ absorption at rest
is due to non-pionic $K^{-}NN$ modes, releasing substantially larger
energy than the corresponding binding energies. A rough estimate for
the residual width, largely due to these modes, is 30-40 MeV. It is
unlikely then that well separated signals could exist in the outgoing
nucleon spectrum of the $(K^{-}, N)$ reaction, except perhaps in the
very special case of kaonic He atoms \cite{AYa99}. Considering the
production cross sections anticipated for such deeply bound $K^{-}$
{\it nuclear} states, we disagree with the estimate made by Kishimoto
\cite{Kis99} for the $(K^{-}, p)$ reaction at 1 GeV/c. Our estimate is 
at the few $\mu$b/sr level only. It would be extremely difficult to 
observe these relatively broad states given such low integrated cross sections.

\section{$\Sigma^-$ ATOMS}
\label{Sigma}

DD analyses of $\Sigma^-$ atoms \cite{BFG94,MFG95} were already reviewed
in the HYP97 meeting \cite{f98}. These analyses suggested
for the first time that the $\Sigma$ nucleus interaction,
which is attractive outside the nucleus, is likely to be repulsive
within. Dabrowski subsequently suggested that such repulsion indeed
is required by the shape of $\Sigma$ quasi-free excitation spectra
\cite{Dab99} (see also his contribution in these proceedings).
Another element that emerged clearly from these works was that the
isovector, Lane potential component of the $\Sigma$ nucleus optical
potential was substantial and of a definite sign. In Ref. \cite{MFG95},
this derivation was assigned to the particularly accurate shift and
width values measured for the atomic $9k$ $\Sigma^-$ level in Pb.
Very recently, Loiseau and Wycech \cite{LWy01} showed that the
relatively accurate $10l$ small width value, derived from the yield
measurement, is instrumental in extracting the value of the $\pi 
\Lambda \Sigma$ coupling constant, assuming that the $\Sigma^-$
absorption occurs peripherally on protons, $\Sigma ^{-}p \to \Lambda
n$, via pion exchange. The value thus deduced, incorporating several
other upper-level widths (of order eV) which are quite justifiedly
evaluated perturbatively, is

\begin{equation}
	{{f^{2}_{\pi \Lambda \Sigma}} \over {4 \pi}} = 0.048 \pm 0.005 \pm
	0.004 \ ,
	\label{equ:pion}
\end{equation}
larger by about 25\% than the value used in recent NSC one boson
exchange models \cite{Stoks99}.

\newpage

\section{$\Xi^-$ ATOMS}
\label{Xi}

Some new information on the $\Xi^{-}$ nucleus interaction has
been recently reported from ($K^-$, $K^+$) counter experiments.
Fukuda et al. \cite{Fuk98} have shown fits to the very low energy part
(including the bound state region) of the $\Xi^{-}$ hypernuclear
spectrum in the $^{12}$C$(K^-, K^+)X$ reaction on a scintillating fiber
active target (KEK experiment E224), resulting in an estimate for 
the depth of the (attractive) $\Xi$ nucleus potential 
($V_{0}^{(\Xi)}$) between 15 to 20 MeV.
The experimental energy resolution of about 10 MeV in this experiment
was too poor to allow identification of any bound state peak structure
which could have given more definitive information on the well depth.
A somewhat cleaner and better resolved spectrum has been recently
presented by the Brookhaven AGS experiment E885 \cite{may00}, suggesting 
that $V_{0}^{(\Xi)} \sim 14$ MeV.

\begin{figure}
\begin{minipage}[t]{80mm}
\epsfig{file=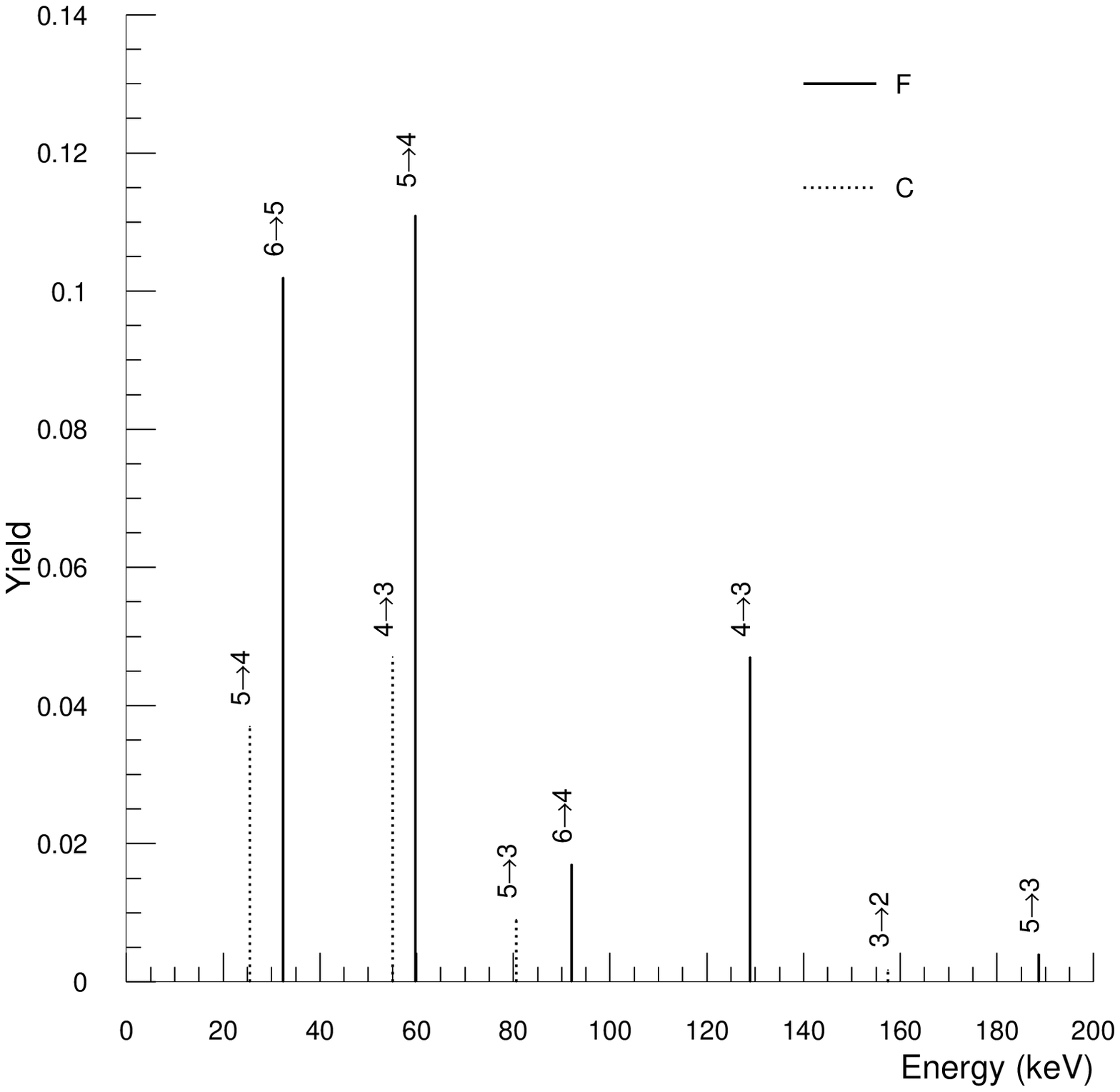,height=87mm,width=80mm,
bbllx=45,bblly=181,bburx=501,bbury=658}
\caption{Calculated $Xi^-$ X-ray spectrum for a teflon target.}
\label{fig:XiCF2}
\end{minipage}
\hspace{\fill}
\begin{minipage}[t]{75mm}
\epsfig{file=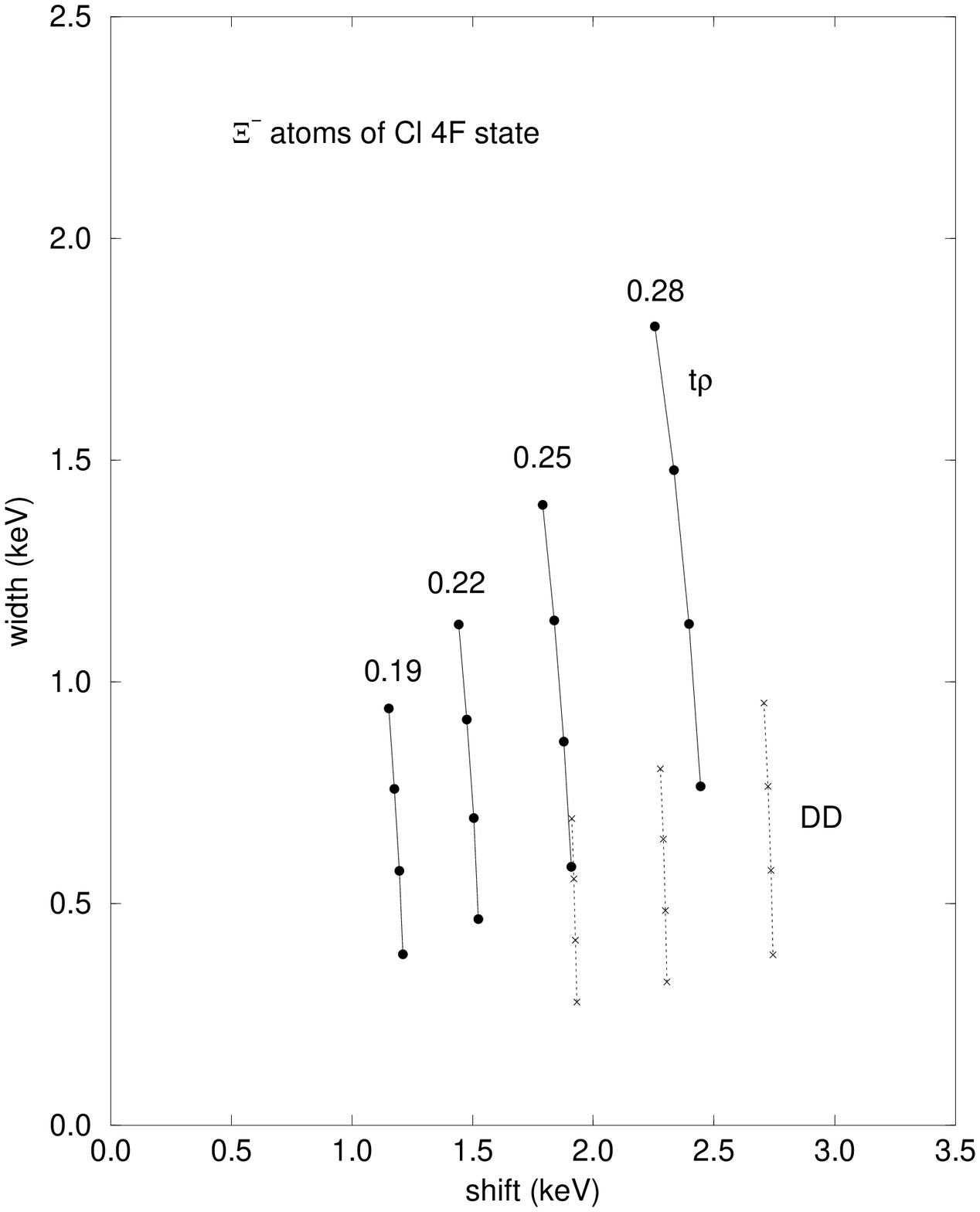,height=87mm,width=75mm,
bbllx=51,bblly=92,bburx=518,bbury=670}
\caption{Calculated shifts and widths for the $4f$ $Xi^-$ atomic state in
Cl for different assumed optical potentials.}
\label{fig:XiCl}
\end{minipage}
\end{figure}

An earlier KEK experiment (E176)
gave evidence for three events of stopped $\Xi^{-}$ in light emulsion
nuclei, each showing a decay into a pair of single $\Lambda$
hypernuclei. The first two events \cite{ABC93} are consistent
energetically with a $\Xi^{-}$ {\it atomic} state in $^{12}$C bound by
$B_{\Xi^{-}}(^{12}{\rm C}) = 0.58 \pm 0.14~~{\rm MeV}$. However, this
value could only be ascribed to capture from the $1s$ state which is
estimated to occur in less than 1\% of the total number of captures.
This binding energy is distinctly larger than the calculated value
$B_{\Xi^-}^{2p}(^{12}{\rm C}){_{_{_{_<}}}\atop ^{^\sim }} 0.32$ MeV
for the $2p$ state,
for a wide range of strong-interaction potentials. Moreover, the $\Xi^-$
capture probability in $^{12}$C from $p$ states is a few percent
at most. The most likely capture in $^{12}$C, as seen in
Fig. \ref{fig:XiCF2} taken from Ref. \cite{BFG99},
occurs from the atomic $3d$ state.
However, the binding of this atomic state (126 keV)
is essentially determined
by the Coulomb potential, and its sensitivity to the
$\Xi$-nucleus strong interaction assumed potential is
of the order of 100 eV,
substantially smaller than a typical error of 100 keV incurred in
emulsion work. We point out that there exist alternative
interpretations of these two events as captures on $^{14}$N, with
binding energies consistent with the calculated value of 175 keV,
for example \cite{ABC93},
$B_{\Xi^{-}}(^{14}{\rm N}) = 0.35 \pm 0.20~~{\rm MeV}$. Furthermore,
a likely interpretation of the third event \cite{NSY97} is due to
capture on $^{16}$O, with
$B_{\Xi^{-}}(^{16}{\rm O}) = 0.31 \pm 0.23~~{\rm MeV}$ (compared to
the calculated value of 231 keV).
Clearly, whereas these emulsion events are consistent with capture from
$3d$ {\it atomic} states, they are useless as a source of
information regarding the $\Xi$-nucleus interaction.

An alternative source of information on the $\Xi$-nucleus
strong interaction would be the measurement of
X-ray energies from transitions between low lying levels of $\Xi^-$
atoms, as suggested recently by Batty et al. \cite {BFG99}. 
The experimental accuracies of the proposed measurements
are sufficient to obtain meaningful information on the $\Xi$-nucleus
interaction.
Full atomic cascade calculations have been performed \cite{BFG99}
for $\Sigma^-$ and $\Xi^-$ atoms and confirmed, as expected,
that the processes within these two hadronic atoms are very similar.
The remaining major differences are in the production reactions.
Whereas $\Sigma^-$ hyperons are produced by
the $p(K^-,\pi^+)\Sigma^-$ reaction at rest, the $p(K^-,K^+)\Xi^-$
reaction occurs at higher energies, thus causing decay losses during
the slowing down time
of the $\Xi^-$ particle to be non-negligible.
Prior to such an experiment it will be necessary to optimize
the experimental setup, which includes a hydrogen production target,
a heavy moderator such as Pb or W, the target to be studied and the
detectors, both for X rays and for the detection of the outgoing
$K^+$, which is essential in order to reduce background.

Adopting an attractive optical potential $V_{\rm opt}$ for the
$\Xi^-$ nucleus strong interaction, and assuming a depth
of 15-20 MeV with an imaginary part of 1-3 MeV,
Batty et al. \cite{BFG99} were able
to propose four targets suitable for X-ray measurements
along the periodic table: F, Cl, I, Pb. A calculated X-ray spectrum for
a teflon (CF$_2$) target is shown in Fig. \ref{fig:XiCF2} and it is seen
that the presence of carbon in the target should not affect the possibility
of observing transitions in $\Xi^-$ F atoms.
This choice of targets also proves to be a sensible one for moderate changes
in $V_{\rm opt}$. Nevertheless, even if the actual potential
turns out to be very different from the one used in these calculations,
only relatively small changes will result,
because the whole phenomenon of hadronic atoms is dominated
by the Coulomb interaction.
Deriving strong-interaction level shifts and
widths for as few as 2 to 4 measured X-ray spectra,
it was shown by Friedman \cite{f98} that $V_{\rm opt}$
may be constrained to a reasonable accuracy over a meaningful
range of nuclear densities. The sensitivity to the $t \rho$ optical
potential assumed in these calculations is demonstrated in Fig.
\ref{fig:XiCl} for Cl. The solid curves connect shift and width
values obtained for fixed values of Re $b_{0}$, listed above the lines.
The four points along each line correspond to values of Im $b_{0}$
from 0.05 fm down to 0.02 fm. Similar results for a DD potential are
also shown.

Extracting $V_{0}^{(\Xi)}$ directly from experiment would provide
extremely valuable information on whether or not strange hadronic
matter \cite{Sch93} exists. Indeed it was found \cite{BGS94} that
once $V_{0}^{(\Xi)}{_{_{_{_>}}} \atop ^{^\sim}}15$ MeV, and for a
strangeness fraction as small as about 0.1 in multi-$\Lambda$
hypernuclei, the free-space strong-interaction conversion $\Xi N \to
\Lambda \Lambda$ becomes Pauli forbidden and the reverse process occurs.
This means that $\Xi$ hyperons together with nucleons and $\Lambda$
hyperons may become stable against strong-interaction baryon
emission, and multistrange nuclei with large strangeness fraction
$-S/A \sim 1$ and small charge fraction $|Q|/A \ll 1/2$, decaying
only weakly, should generallly exist; for a recent update see
Ref. \cite{SBG00}.

\section*{ACKNOWLEDGEMENTS}
\label{sec:acknow}

I would like to acknowledge a longstanding collaboration on the topics here
reviewed with Drs. C.J. Batty, A. Cieply, E. Friedman and J. Mare{\v s}, as 
well as useful discussions with Drs. A. Ramos and J. Schaffner-Bielich.
This work was partly supported by the trilateral DFG contract GR 243/51-2.

\end{document}